%% file: sn-article.tex
\documentclass[pdflatex,sn-mathphys]{sn-jnl}

\usepackage{subfiles}

\jyear{2021}%

\theoremstyle{thmstyleone}%
\theoremstyle{thmstyletwo}%

\theoremstyle{thmstylethree}%

\begin{document}

\title[Article Title]{Target specific peptide design using latent space approximate trajectory collector}


\author[3,4]{\fnm{Tong} \sur{Lin}}
\equalcont{These authors contributed equally to this work.}

\author[1]{\fnm{Sijie} \sur{Chen}}
\equalcont{These authors contributed equally to this work.}

\author[2]{\fnm{Ruchira} \sur{Basu}}

\author*[2]{\fnm{Dehua} \sur{Pei}}\email{pei.3@osu.edu}

\author*[1]{\fnm{Xiaolin} \sur{Cheng}}\email{cheng.1302@osu.edu}

\author*[3]{\fnm{Levent Burak} \sur{Kara}}\email{lkara@cmu.edu}

\affil*[1]{\orgdiv{College of Pharmacy}, \orgname{The Ohio State University}, \orgaddress{\street{281 W Lane Ave}, \city{Columbus}, \postcode{43210}, \state{OH}, \country{United States}}}
\affil[2]{\orgdiv{Department of Chemistry and Biochemistry}, \orgname{The Ohio State University}, \orgaddress{\street{281 W Lane Ave}, \city{Columbus}, \postcode{43210}, \state{OH}, \country{United States}}}

\affil[3]{\orgdiv{Mechanical Engineering Department}, \orgname{Carnegie Mellon University}, \orgaddress{\street{5000 Forbes Ave}, \city{Pittsburgh}, \postcode{15213}, \state{PA}, \country{United States}}}
\affil[4]{\orgdiv{Machine Learning Department}, \orgname{Carnegie Mellon University}, \orgaddress{\street{5000 Forbes Ave}, \city{Pittsburgh}, \postcode{15213}, \state{PA}, \country{United States}}}
\affil[5]{\orgdiv{Electrical and Computer Engineering Department}, \orgname{Carnegie Mellon University}, \orgaddress{\street{5000 Forbes Ave}, \city{Pittsburgh}, \postcode{15213}, \state{PA}, \country{United States}}}

\abstract{Despite the prevalence and many successes of deep learning applications in de novo molecular design, the problem of peptide generation targeting specific proteins remains unsolved. A main barrier for this is the scarcity of the high-quality training data. To tackle the issue, we propose a novel machine learning based peptide design architecture, called Latent Space Approximate Trajectory Collector (LSATC). It consists of a series of samplers on an optimization trajectory on a highly non-convex energy landscape that approximates the distributions of peptides with desired properties in a latent space. The process involves little human intervention and can be implemented in an end-to-end manner. We demonstrate the model by the design of peptide extensions targeting Beta-catenin, a key nuclear effector protein involved in canonical Wnt signalling. When compared with a random sampler, LSATC can sample peptides with $36\%$ lower binding scores in a $16$ times smaller interquartile range (IQR) and $284\%$ less hydrophobicity with a $1.4$ times smaller IQR. LSATC also largely outperforms other common generative models. Finally, we utilized a clustering algorithm to select 4 peptides from the 100 LSATC designed peptides for experimental validation. The result confirms that all the four peptides extended by LSATC show improved Beta-catenin binding by at least $20.0\%$, and two of the peptides show a $3$ fold increase in binding affinity as compared to the base peptide.}



\maketitle

\subfile{contents/1_introduction}

\subfile{contents/2_result}

\subfile{contents/3_discussion}

\subfile{contents/4_method}


\bibliography{sn-bibliography}


\end{document}

%% file: contents/1_introduction.tex
\section{Introduction}\label{sec1}
Therapeutic peptides are a class of pharmaceutical agents that are distinct from small molecule drugs due to their unique biochemical and therapeutic characteristics. In recent years, many peptide drugs have been found to have superior potency and safety profiles than small molecule drugs \cite{cite1}. Peptides can disrupt unwanted protein-protein interactions (PPI) that have often been implicated to play a role in cancer development and progression. One such example is the deregulation of Wnt/beta-catenin/T-cell factor (Tcf) signaling common in many human cancers. Thus, the development of peptide-based PPI inhibitors has become one of the most topical directions in cancer drug research.


Structure-based design of peptide inhibitors for a specific protein target has long been an empirical task. Traditionally, peptide design has been focused on sequence perturbation, including residue mutation, interchain residue exchange, alanine scanning and chemical modification \cite{cite2_1,cite2_2}, which is guided by structural information obtained for the protein or protein-peptide complex system. The key limitation of this strategy is the negligence of potential secondary structure changes upon sequence perturbation and how the resulting changes may shift the protein-peptide binding structures. High-Throughput-Screening (HTS) has also been widely applied in peptide design. HTS is a brute force method to identify bioactive peptides by rapidly conducting thousands to millions of biochemical, genetic, or pharmacological assays. However, HTS demands highly specialized instrumentation, development and adoption of appropriate bioassays, and high quality peptide libraries. Additionally, actives discovered in HTS are often serendipitous \cite{cite3}.  To complement HTS, virtual screening, such as peptide docking, has been heavily used for peptide inhibitor design. \cite{cite4}. Peptide docking is usually composed of a sampling technique to explore peptide conformations and a scoring function to evaluate the strength of peptide binding for all sampled peptide binding poses. The two-step protocol takes at least minutes to evaluate one protein-peptide complex \cite{cite5}, which limits its ability in sequence exploration and thus its use in de novo peptide design.

In recent years, machine learning-based molecular design has witnessed rapid development. A very important area in deep generative models is the efficient representation of molecules. Prior to the deep learning era, fingerprint descriptors such as Morgan fingerprints \cite{cite6} for small molecules and atom-pair fingerprints for proteins \cite{cite7} are prevalent. However, these representations that encode the chemical and structural information of individual molecules are not task specific \cite{cite8}. Deep learning models have been developed to address this limitation by learning a unified representation in a large dataset and then fine tuning this representation for a specific task. During the past few years, the research on molecular representation has been shifted to string transformation, most of which has borrowed the idea from natural language processing. There are two major types of model framework. The first is the recurrent neural network, such as Long-Short Term Memory (LSTM)\cite{cite9}, Gated Recurrent Units \cite{cite10} and Recurrent Attention \cite{cite11}. These models have been utilized to predict molecular properties, such as solubility, toxicity \cite{cite12,cite13,cite14}. The other type is transformer \cite{cite15} based models, which incorporate a multi-head attention mechanism to process sequential data more efficiently. The transformer based architectures such as ProteinBert\cite{cite16} and ProteinTrans\cite{cite17} have been frequently used in protein representation, and have shown great success in multiple protein downstream classification or regression tasks (e.g. secondary structure classification, fluorescence prediction). Recently, the graph representation of molecules has gained great attention due to the graph's ability to include more detailed structural and spatial information \cite{cite18}. Graph neural network has been applied to process molecular graphs and perform property prediction \cite{cite19,cite20,cite21}.


Although the learned molecular representation has been exploited in many property prediction models, few models are about molecular generation. In 2017, Bjerrum used the RNN network to generate valid molecules \cite{cite22}. Since then, several studies have been published on the generation of valid molecules with optimized general properties such as logP, TPSA and QED \cite{cite23,cite24,cite25}, while work on protein sequence design has been scarce, and most of it has focused on a single protein's general property design such as the length, the stability and the isoelectronic point\cite{cite26,cite27}. Drug design targeting protein interaction has been less explored. In 2020, Das proposed a method for antimicrobial drug design using rejection sampling to search appropriate molecular SMILES in a latent space\cite{cite28}. In 2022, Castro proposed a gradient based latent space search method for designing a protein sequence against the third complementarity-determining region of the ranibizumab antibody heavy chain \cite{cite29}. The two papers are most relevant to our work, albeit the first work is not on peptide drug design and the second limits its usage to one protein that has an existing dataset of 60,000 samples. The main challenge in protein-specific peptide design is the inadequacy of accurate binding affinity data due to the large computational or experimental cost and the immense peptide space. To our knowledge, deep learning models for protein specific peptide generation don't yet exist.

\begin{figure}[h]
\centering
    \vspace{-5pt}
    \includegraphics[width=0.8\textwidth]{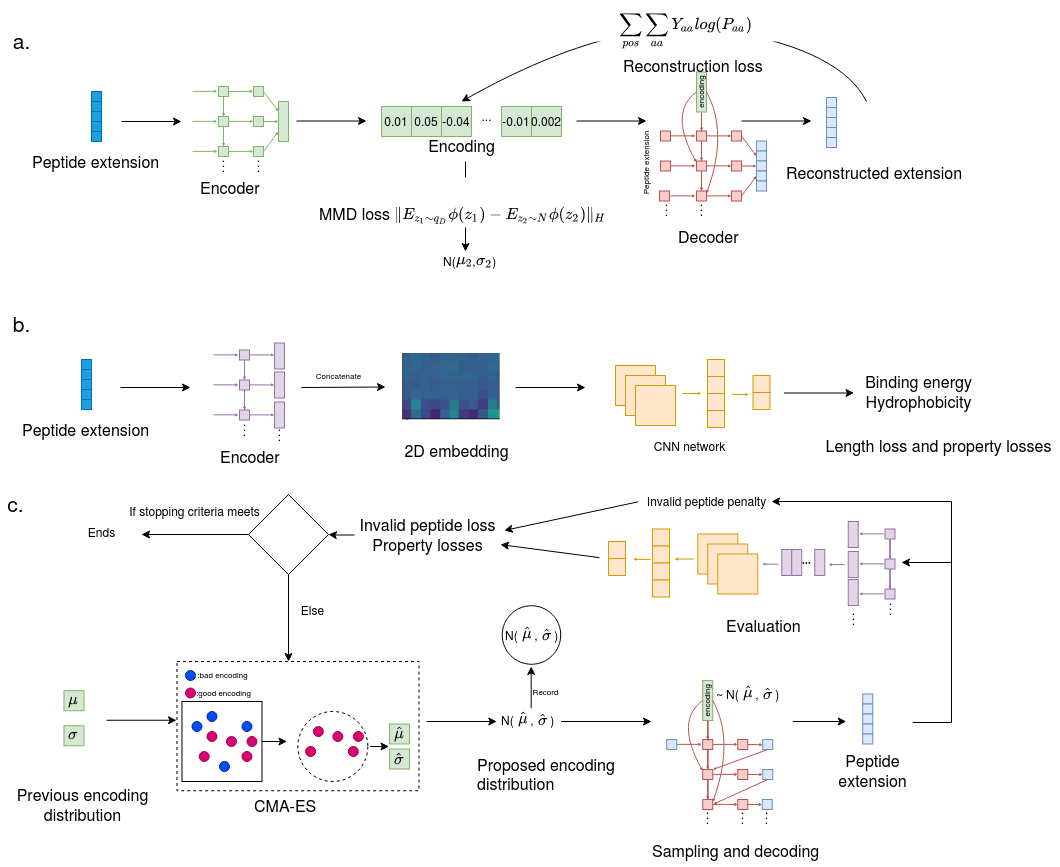}
    \caption {(a). Peptide extension reconstruction network. A reconstruction loss is used. The distribution of the encoding is regularized to a Gaussian distribution.(b). Property prediction network. The model estimates the hydrophobicity and the binding energy of a peptide extension. (c). Optimization process for generative model. A Gaussian sampler in the latent space is learned using CMA-ES. Reconstruction network is used in decoding. Property prediction network is used in the evaluation as a surrogate model of Pyrosetta. The samplers are recorded in each of the iteration. The total loss includes the desired peptide properties and penalty to sample invalid encodings.}
    \vspace{-5pt}
    \label{fig:model_figure}
\end{figure}

In this paper, we propose a novel protein specific peptide generation scheme (Figure \ref{fig:model_figure}), called Latent Space Approximate Trajectory Collector (LSATC). We implement a GRU based Wasserstein auto encoder (WAE) \cite{cite31} to obtain 1D continuous latent space representation of peptides, as shown in Figure \ref{fig:model_figure}(a). To circumvent the data scarcity problem, we design a feedback loop using CMA-ES \cite{cite32} to optimize the generator in a peptide latent space and collect the generator's trajectory. The process is illustrated in Figure \ref{fig:model_figure}(c). A fast feedback evaluator is crucial to the scheme. Since the current binding evaluator (e.g. docking) is computationally costly, we train an efficient surrogate model to boost the evaluation time from minutes to microseconds (Figure \ref{fig:model_figure}(b)). For the surrogate model training, the binding energies and the hydrophobicities of a reasonably small number of randomly sampled peptides are evaluated using Pyrosetta and Biopython, respectively. Finally, we sample peptides on the explored trajectory whose associated losses are lower than a predefined threshold value. We note that the increase in speed offered by the surrogate model enable us to explore the peptide encoder space for peptide inhibitor design and optimization. Additionally, the incorporation of a biophysics-based model in our deep learning can facilitate peptide candidate selection during the post processing stage.
 
Here, we test our LSATC model in a multi-objective peptide extension task. Specifically, we aim to improve the binding of a base peptide "YPEDILDKHLQRVIL" with beta-catenin by extending its N-terminus, and to reduce the hydrophobicity of peptides to minimize non-specific binding. For simplicity, we only consider peptide extensions of 5 amino acids, which limits the search space to 3.2 million peptides. However, it is worth noting that LSATC is not limited to generation of fixed length peptides. The entire peptide extension process, including dataset generation and model training, takes two days to finish. Our LSATC model is proven to be much more efficient in generating desired peptides than random generation and several other commonly used generative models. The \emph{in vitro} results show that our generated peptides are not only less hydrophobic but also more potent than the experimental baseline result, with the highest improvement of 3 fold. 

Our contributions in the paper is as following, 
\begin{itemize}
 \item We design a peptide generator that can efficiently discover more novel protein-specific peptide sequences when no or very few binding data exists.
 \item We give insights of how and why our machine learning based peptide generator works under the condition that the data is scarce. This increases the interpretability of the model.  
 \item Our proposed design pipeline is \emph{in vitro} test ready and highly automated. We create a peptide filtering strategy to select desired number of peptides among the sampled high quality peptides for \emph{vitro test}. This erases the needs of experimentalists' manual inspection to finalize the selected testing peptides. 
 \end{itemize}

%% file: contents/2_result.tex
\section{Results}\label{sec2}
In \emph{in silico} evaluation, LSATC can much more efficiently generates peptide extensions with low binding energy and low hydrophobicity than the random generation, Gaussian mixture model (GMM) and the conditional Wasserstein Autoencoder(cWAE). In \emph{in vitro} test, all the sampled peptide extensions largely improve the base peptide binding score. In this section, we will present the analysis and the results of each component of the LSATC.

\subsection{Dataset preparation}
In LSATC, the sequence reconstruction model is used for converting the amino acid representation from discrete letters to continuous numbers. The property surrogate model is used for a fast peptide binding energy and hydrophobicity evaluation. A 500,000 unlabelled peptide extension dataset and a 50,000 labelled peptide extension dataset are prepared for the training of the sequence reconstruction and the surrogate model, respectively. For both datasets, all the peptide extensions are unique and randomly generated.

\begin{figure}[h]
\centering
    \vspace{-5pt}
    \includegraphics[width=0.9\textwidth]{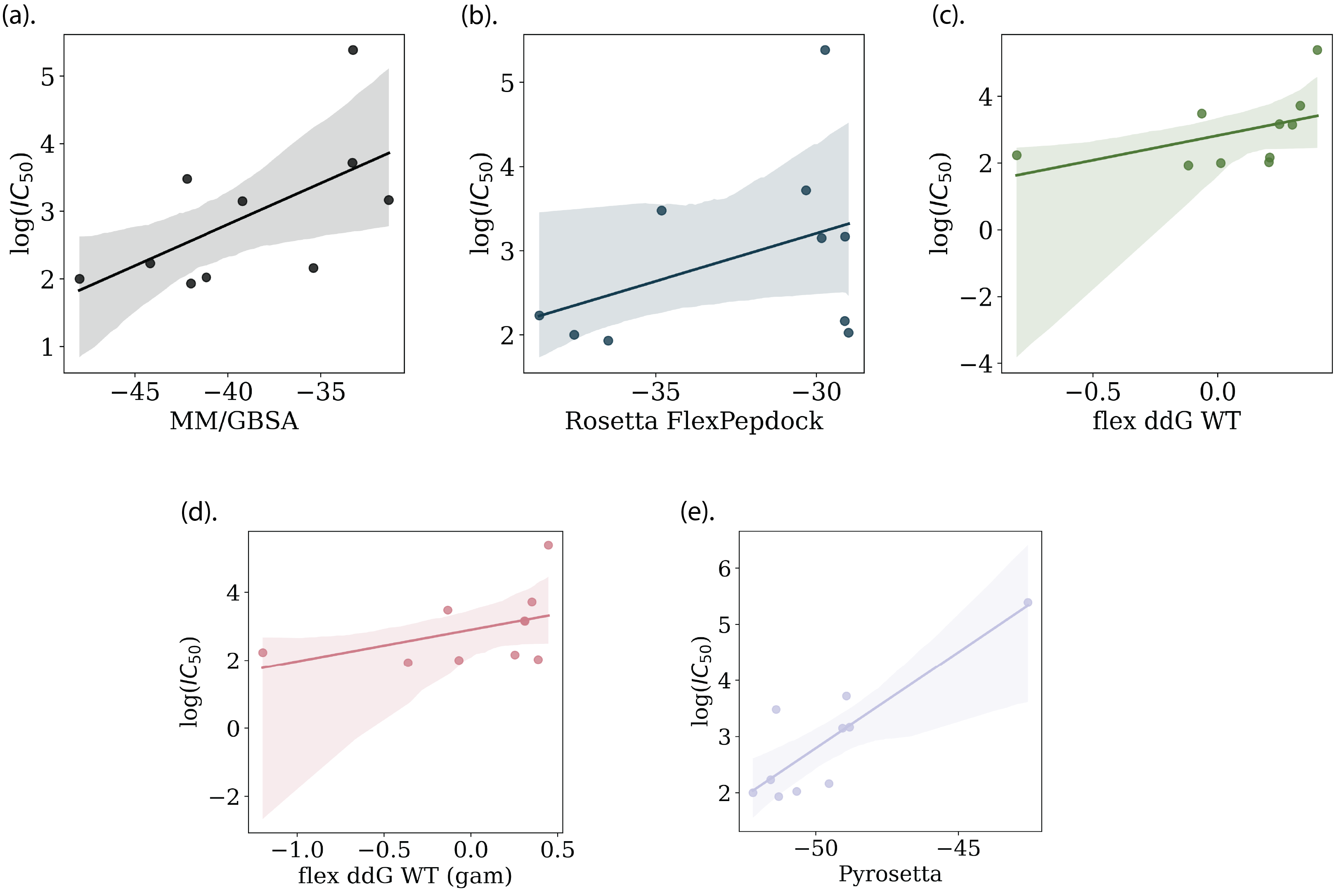}
    \caption {(a). The regression plot of the log of in vitro IC50 and binding energies calculated by MM/GBSA. (b). The regression plot of the log of in vitro IC50 and binding energies calculated by Rosetta FlexPepDock. (c). The regression plot of the log of in vitro IC50 and binding energies calculated by flex ddG. (d).  The regression plot of the log of in vitro IC50 and binding energies calculated by flex ddG(gam). e). The regression plot of the log of in vitro IC50 and binding energies calculated by Pyrosetta.}
    \vspace{-5pt}
    \label{fig:binding_tool_cmp}
\end{figure}

Each peptide in the labelled dataset possesses two properties - hydrophobicity and beta-catenin binding score. The hydrophobicity is a quantity to show the tendency of water to exclude nonpolar molecules. It is a sequence dependent property. We calculate the hydrophobicity using Biopython. The Kyte-Doolittle \cite{KD_scale} scale is used for measuring the degree of the hydrophobicity for each amino acids. We select a window size 3 and calcute the moving averages by sliding the window on the peptides. The hydrophobicity is computed by summing all the movin averages. Note that we use fixed length of the peptides. Thus, the hydrophobicity does not need to be normalized with respect to the peptide lengths. The binding score is calculated according to \cite{binding_calc}. It is a weighted sum of the Van Der Waals force energies, solvation energy, residue–residue pair potentials, hydrogen bond energies, electrostatics energy and internal energy of sidechain rotamers. The unit for the energy is $\frac{kcal}{mol}$. The distance threshold to define interacting atoms is $10\AA$. The computation of binding scores requires 3D structural information of the peptide-protein complexes. Such information is obtained through mutational substitution of an initial peptide-protein complex structure, for which a detailed description is given in Section 4.1. We have tested 5 different methods for binding energy calculation, MM/GBSA, Rosetta FlexPepDock, flex ddG, flex ddG(gam) and Pyrosetta. In Figure \ref{fig:binding_tool_cmp}, we plot the correlations between the experimental binding data and those estimated from the five methods for 10 assayed peptides. The result shows Pyrosetta has the best linear correlation in lower energy regions. MM/GBSA and Rosetta FlexPepdock score peptides in the whole region. Thus, we select Pyrosetta to compute the binding energies of the 50,000 peptides. Although Pyrosetta is a lightweight software package, the generation of the labelled dataset still takes around 12 hours to finish. 

To properly train the surrogate model, the log transformation is performed to normalize the binding energy to reduce the outliers' effect. The detail of the dataset generation process is shown in Section 4.2.     

\begin{figure}[h]
\centering
    \vspace{-5pt}
    \includegraphics[width=0.8\textwidth]{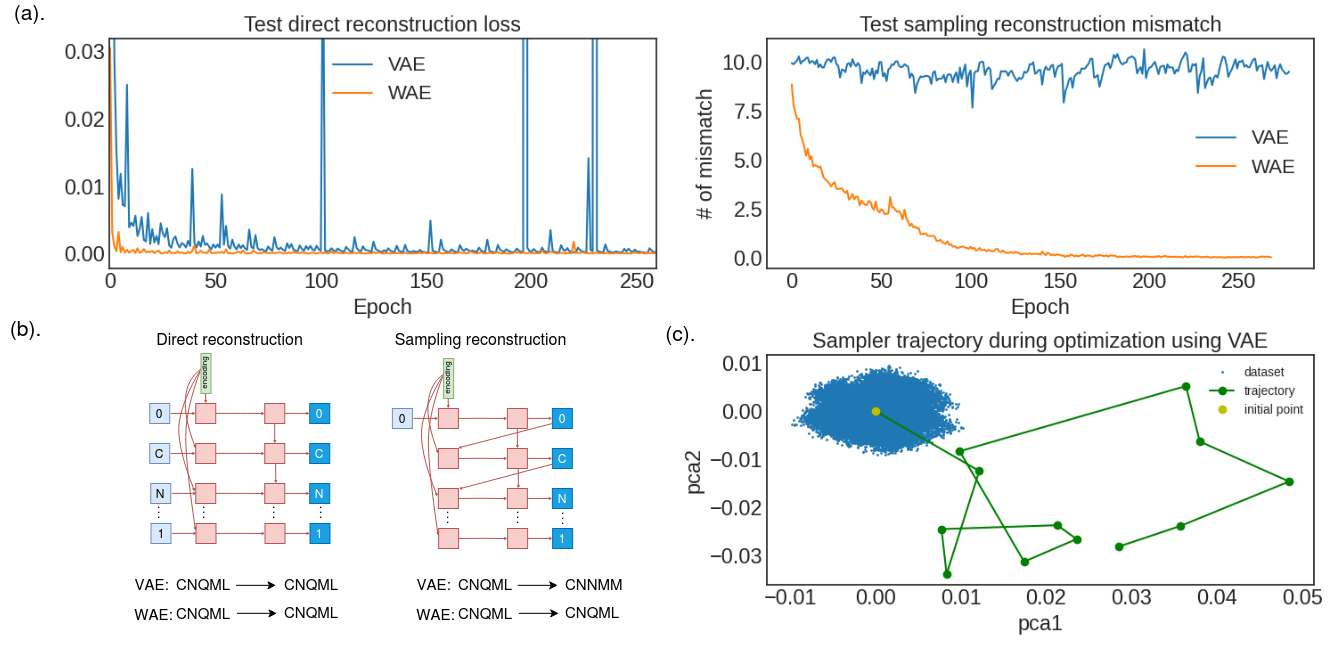}
    \caption {(a). The direct reconstruction loss (left) and the sampling reconstruction loss (right) during the training.(b). A illustration of the different between the direct reconstruction and the sampling reconstruction. Two results of their reconstruction results on VAE and WAE are shown. (c). An example of the optimization trajectory of the sampler for a decoder with low sampling reconstruction accuracy. The sampler is optimizer in a different latent space than the latent space described by the encoder which is represented by the dataset here.
    }
    \vspace{-5pt}
    \label{fig:WAE_losses}
\end{figure}

\subsection{Sequences reconstruction}
The mapping between the peptides represented by the amino acids and their properties is not smooth because the amino acid space is discrete. This often leads to the ineffectiveness of an optimizer to find better peptides. To alleviate the problem, a sequences reconstruction model is created to represent peptide sequences in continuous number and to decode the number back to the original amino acid space. To achieve the goal, we use a gated recurrent unit (GRU) generative auto encoder framework for sequence reconstruction. The sequence reconstruction process during the training phase and the inference phase are different. We call the first direct reconstruction and the second sampling reconstruction. We show the difference of the two in Figure \ref{fig:WAE_losses}(b). When the decoding is performed, the known encoding and the amino acid at the previous position needs to be used to output the current amino acid. For the direct reconstruction, the previous amino acid is always correctly inputted because it is known during the training; however, for the sampling reconstruction, it has to use the predicted previous amino acid as an input to infer the current amino acid. This is because the true previous amino acid is not accessible during the inference phase. Sampling reconstruction is important for our model. A deficient reconstruction model with low sampling reconstruction accuracy could result in an encoding shift during our later optimization stage. In Figure \ref{fig:WAE_losses}(c), we show an optimization trajectory of an encoding sampler using a deficient model. The optimization trajectory is the trajectory of the encoding during the search for a good encoding, which we will illustrate in Section 2.4. Due to the discrepancy between the encoder space and the decoder space, the optimizer, which needs feedback from decoded sequences, does not search in or near the encoder's output space represented by the dataset. 

In this study, we have compared a commonly used variational AE (VAE) and a Warseetein AE (WAE) in reconstruction. VAE uses KL divergence to regulate the encoding distribution to be the target distribution while WAE uses the Wasserstein distance to achieve the goal. The details of the two models are described in Section 4.3. The right plot in Figure \ref{fig:WAE_losses}(a) shows the test mean square error loss for the direct reconstruction during the training. We find that both models have low test direct reconstruction loss. On the right figure, we show the sampling reconstruction error as the number of the mismatch between the input and the reconstructed output. The sampling reconstruction error of the VAE model stays high while the WAE manages to reduce the error close to 0 after 150 epochs even though both models have low direct reconstruction loss. In fact, the VAE model tends to repeat the previous input as the output during the inference stage. An example is shown in Figure \ref{fig:WAE_losses}(b). Based on the above experiments, we choose a GRU-based WAE model to encode and decode the peptide sequences as it has good results of both direct reconstruction and sampling reconstruction.   

\begin{figure}[h]
\centering
    \vspace{-5pt}
    \includegraphics[width=0.8\textwidth]{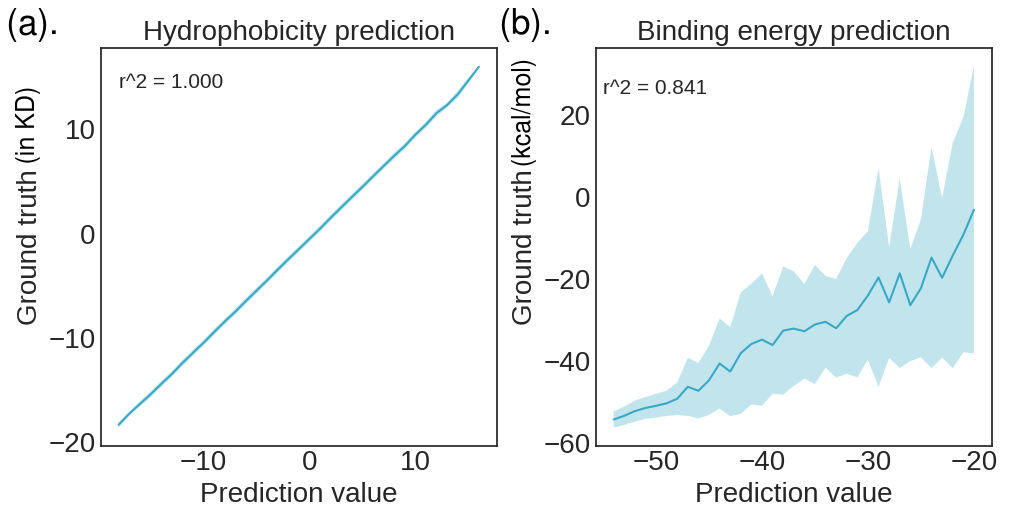}
    \caption {(a). The $r^{2}$ plot of the hydrophobicity. The $r^{2}$ value is almost 1. (b). The binding score prediction plot. The $r^{2}$ value is calculated between the prediction mean in a small interval and the mean of the true values whose associated predictions fall into the interval. The uncertainty represents the range of the ground truth at each predicted value. It is apparent that the smaller the prediction value is, the smaller the uncertainty is. 
    }
    \vspace{-5pt}
    \label{fig:surrogate_losses}
\end{figure}

\subsection{Surrogate model prediction}
Accurate evaluation of peptide-protein binding free energies is computationally demanding due in a large part to the considerable conformational, translational, and rotational changes underlying the binding process that are difficult to sample. Hence, to be incorporated into an iterative optimization process a fast but accurate free energy evaluation method is required. To this end, we train a surrogate model to predict the beta-catenin binding scores and the hydrophobicity of peptides on our labelled dataset.  



The prediction network is similar to the reconstruction network except that a convolution neural network (CNN) is added at the end to predict the binding energy and the hydrophobicity. We have tested several different machine learning models, among which the CNN model achieves the lowest test mean square error in predicting the hydrophobicity and the binding energy. All of the test results are shown in Support Information. 

The predicted hydrophobicity shows excellent correlation with the experimental data (Figure \ref{fig:surrogate_losses}(a)). This suggests that our surrogate model can accurately capture the sequence level property of peptides.

The binding score turns out more difficult to predict as the surrogate model relies on peptide sequences exclusively. It can be considered as a deep learning-based quantitative structure activity relationship (DL-QSAR) model. To balance the speed and the accuracy, our goal is not to obtain predictions highly correlated with the true activity values, but to be able to rank the order of peptide binding so that our focus can be put on those peptides with a low binding score. What we need is that when Pyrosetta predicts a low binding score for a peptide, the surrogate model would also predict a low binding score for the peptide. The surrogate model shows a reasonably good correlation between the predicted and Pyrosetta calculated values with r2 of 0.841. Importantly, the uncertainty decreases as the binding score decreases as shown in \ref{fig:surrogate_losses}(b). This indicates that a peptide is more likely to have a true low score if it is predicted to have a low score by the surrogate model, which is well suited for our optimization need.


\subsection{CMA-ES optimization results}
\begin{figure}[h]
\centering
    \vspace{-5pt}
    \includegraphics[width=0.9\textwidth]{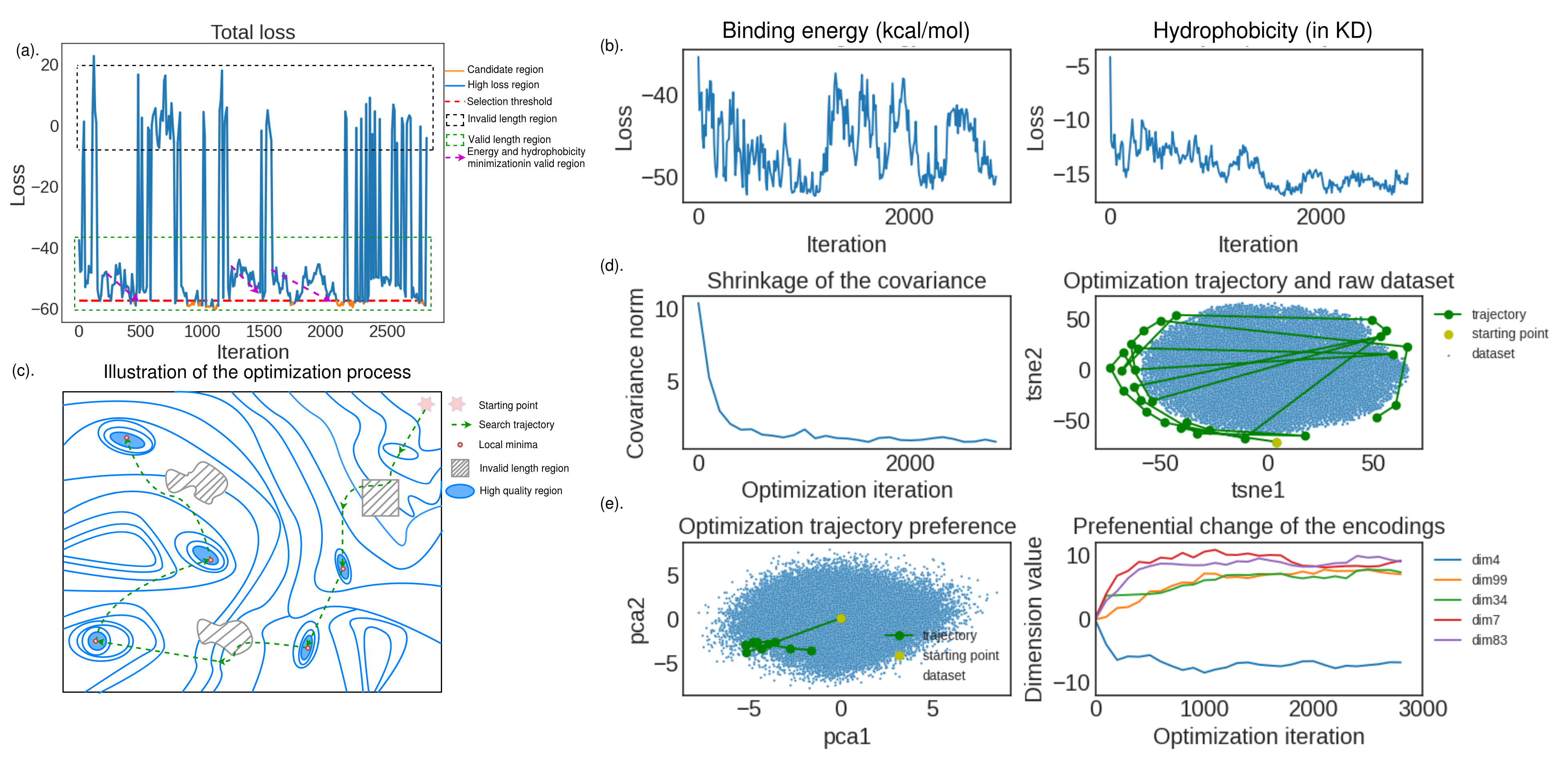}
    \caption {(a). The CMA-ES training loss. There is a high loss region caused by the invalid peptide penalty and a low loss region representing a rich region of valid peptides. In the low loss region, the primary goal is to reduce the property losses. (b). The averaged binding scores during the optimization. (left). The averaged hydrophobicity during the optimization(right). (c). An illustration of the sampler's behavior during the optimization (d). The magnitude of the covariance during the optimization. The magnitude is calculated as the sum of the absolute value of each element (left). A 2d-tnse plot to show the trajectory of the sampler mean. The sampler is searching around the periphery of the labelled dataset (right). (e). A 2d-pca plot to show the preferential direction of the encoding change. It is obvious that most of the search happens in the bottom left corner on the edge of the labelled dataset (left). The value change for 5 of the encoding dimensions.They are tuned either to positive or negative direction, indicating the tuner has a tuning preference (right).}
    \vspace{-5pt}
    \label{fig:CMA_ES_losses}
\end{figure}

CMA-ES is designed to optimize a Gaussian distribution's mean and variance in a gradient-free approach. We parameterize the region of the good encodings by a Gaussian distribution which we call an encoding sampler. The space of the properties w.r.t the encoding is highly non-convex. Thus, the CMA-ES is a good optimizer for optimizing this sampler model. Specifically, we design the sampler as an isotropic Gaussian distribution $N(\mu,\sigma)$ that can sample good peptides in a latent space, where the dimension of the mean $\mu$ is the same as that of the encodings. $\mu$ and $\sigma$ need to be optimized so that the sampled sequences have desired properties of both binding energy and hydrophobicity. We implement CMA-ES for this purpose. The loss function of the CMA-ES is comprised of three components: the binding energy and the hydrophobicity and the penalty for invalid peptides. The penalty is applied to help the sampler quickly escape from an invalid peptide rich region, which we define as more than $80\%$ of the sampled peptides being invalid. The details of the loss design are presented in Section 4.4. Figure \ref{fig:CMA_ES_losses}(a) shows the losses change during the optimization. The high losses enclosed by a black box indicate that the sampler enters an invalid peptide rich region. When moving into a valid peptide rich region enclosed by a green box, the sampler has a strong tendency to generate peptides of low binding energy and low hydrophobicity as shown by the pink arrow. The right plot in Figure \ref{fig:CMA_ES_losses}(d) shows a trajectory of $\mu$ during the optimization along with the encodings of the labelled dataset. $\mu$ appears to move around the periphery of the labelled dataset. Such behavior balances the novelty and the similarity to the labelled dataset. In the left plot of Figure \ref{fig:CMA_ES_losses}(d), the learned $\sigma$ converges to small values as the optimization process goes. This behavior of the sampler's exploration process can, thus, be interpreted as the following. When the sampler reaches an invalid peptide rich region, it quickly escapes from the region. However, when the sampler enters a valid peptide rich region, it starts performing local optimization on the binding score and the hydrophobicty via searching in a small region around $\mu$. Such two processes alternate throughout the optimization. This is illustrated in \ref{fig:CMA_ES_losses}(c). Concentrating on a small region in the latent space is reasonable as it can be seen from the left plot in Figure \ref{fig:samler_cmp}(a) that the total score w.r.t the encoding is highly non-convex. If the search region is large, it is easy to sample low quality peptides.

Figure \ref{fig:CMA_ES_losses}(b) shows the trajectories of binding score and hydrophobicity during optimization. The minimum hydrophobicity value is around -15 in KD scale and the minimum binding score is around -50$\frac{kcal}{mol}$, all calculated from generated unseen peptides. Note that in the labelled dataset, the range of the binding energy is from -59 $\frac{kcal}{mol}$ to 4782$\frac{kcal}{mol}$ with standard deviation of 220$\frac{kcal}{mol}$. The range of the hydrophobicity is from -19 to 17 with stadnard deviation of 6 in KD scale. Assuming that the surrogate model can approximate the two properties well, both minimum values are close to the lowest values in the labelled dataset. Thus, it is important to collect the sampler information at these local minima. We further explains the sampler collections in Section 2.5.

During the optimization, we find that the optimizer has a preference in tuning $\mu$ to a certain direction. The left plot in Figure  \ref{fig:CMA_ES_losses}(e) shows that the $\mu$'s trajectory prefers high negative values on the first and second principal components. The right plot in Figure \ref{fig:CMA_ES_losses}(e) shows that the top 5 encoding dimensions whose values change the most during the optimization. It is obvious that these values change in a preferential direction during the optimization. The interpretability of high dimensional encodings for any reconstruction models is a long-standing research problem, which causes trouble in tuning the encodings to generate objects with specific properties. The CMA-ES optimizer proposed here provides a way to auto-tune the encodings to accomplish property specific generation tasks.  

CMA-ES, as a gradient free method, is ideal for peptide generation. The reason is three fold. First, the encodings are regularized to be a Gaussian distribution in the sequence reconstruction model. This aligns with the CMA-ES assumption where the sampler is also a Gaussian. Second, CMA-ES is a gradient-free method. It is not guaranteed that any generated encoding can be decoded into a valid sequence. Thus, an important objective is to make sure that the generated encodings are valid (e.g. the decoded sequence has a length of 5). There is no gradient information about this objective w.r.t the encodings. Third, the loss w.r.t the encoding is likely to be highly non-convex. Many local minima could exist and we would like to collect the information in these local minima. It is easy for a gradient based method to be stuck in a local minimum while a gradient free method can climb over the barriers between these minima.

\subsection{Sequences sampling results}
\begin{figure}[h]
\centering
    \vspace{-5pt}
    \includegraphics[width=0.9\textwidth]{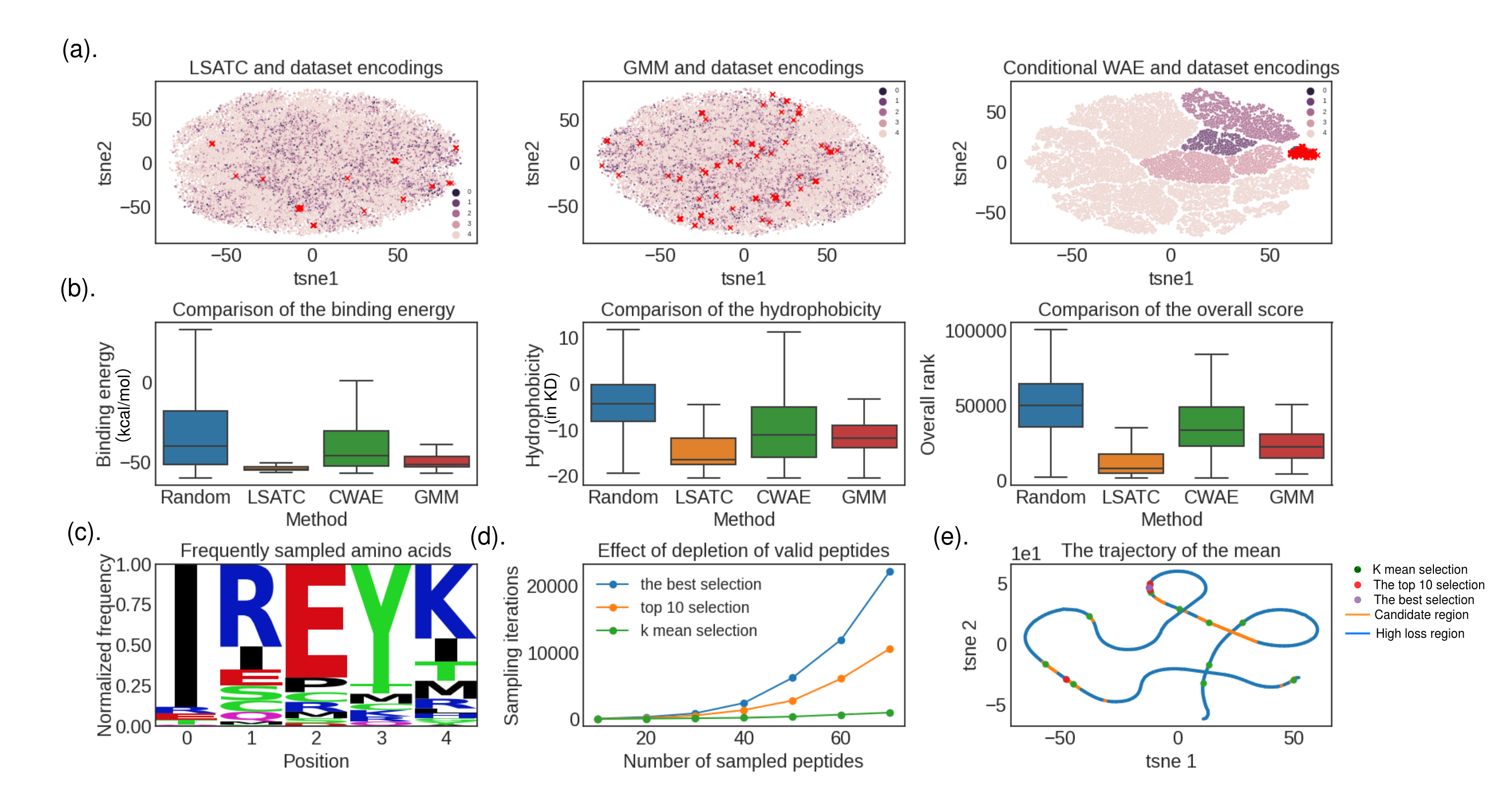}
    \caption {(a). 2d-tsne plot to show how the samples from different generative models positioned in the training high quality peptides. The models are LSATC(left), GMM(mid) and CWAE(right).  
     (b). The comparison of the sampled peptide qualities from random selection, LSATC, CWAE and GMM. Note that it is necessary to have an overall score to describe a peptide as the binding plot and the hydrophobicity plot losses the peptide specificity. (c). A logo plot to show the most sampled amino acids at each extension position. (d). The depletion of the valid encoding. Sampling batch is fixed for each iteration. It is obvious that the required iteration rapidly increases as the number of the request peptides increases except K-mean selection. (e). The optimization trajectory for the sampler and the final selected samplers using different selection strategy.}
    \vspace{-5pt}
    \label{fig:samler_cmp}
\end{figure}

As shown in Figure \ref{fig:CMA_ES_losses}(d), the $\sigma$ tends to be small during the optimization. In our experiment, we find that sampling valid encodings at a single ($\mu,\sigma$) becomes harder as the sampling iteration increases. The small $\sigma$ ensures the quality of the sampled peptides, but suffers from the depletion of valid encodings. We iteratively collect the valid encodings until reaching a desired number. At a single location the required sampling iteration is found to rapidly increases as the requested number of peptides increases. This is evident from Figure \ref{fig:samler_cmp}(d) for the best selection curve, which corresponds to the sampling at a single location. Thus, the peptide depletion problem could be alleviated by collecting peptides from multiple regions, whereas selecting the appropriate regions could be tricky. For example, the sampler could linger around the same region for several iterations and produce similar property losses. In this case, some of the collected peptides with the lowest losses are likely from the same region, which would still suffer from the encodings depletion problem. This situation is shown as the red dots in Figure \ref{fig:samler_cmp}(e). To solve this problem, we collect the sampler trajectories ($\mu$,$\sigma$) whose losses are in top 500, which are referred to as the candidate trajectories. These regions are shown in orange in Figure \ref{fig:samler_cmp}(e) and their corresponding losses are shown in orange in Figure \ref{fig:CMA_ES_losses}(a). To obtain distant samplers, we implement K mean algorithms to cluster the 500 samplers and select the representative samplers from individual clusters. The selected samplers are shown as green dots in Figure \ref{fig:samler_cmp}(e), which are evidently better separated than the red dots. 

Our proposed LSATC is a collection of high quality samplers each corresponding to a distant small region in a local minimum. To see if such a strategy is effective, we compare it with algorithms that directly model known high quality encodings. Since hydrophobic peptides tend to bind proteins non-specifically, we define an overall score to unify the hydrophobicity and the binding score. We rank the binding energy and the hydrophobicity of a peptide separately and then calculate the overall score as the summation of the two. The lower the score, the better the quality. To select peptides for training, we divide peptides in the labelled dataset into 5 classes. The label 0 peptides have an overall score below 10,000. These peptides have the highest quality and account for 456 out of the 50,000 peptides. We implemented two models on this dataset. The first one is a Gaussian mixture model (GMM), which is used to model the encoding distributions of the label 0 peptides. The granularity of the distribution approximated by GMM depends on the cluster number. The more clusters, the finer the distribution is. However, the model is more likely to suffer from the depletion of valid encodings as the cluster number becomes larger. This is because each of the distribution modes can be very concentrated as there are only 456 encodings in the dataset. We show in Support Information that 200 clusters give the best result. The second model is a conditional WAE model, which manually separates the encodings of each class by concatenating the label into the encoding. It models the distribution of each of the labelled encodings separately. The high quality peptides can be directly sampled by concatenating the label 0 and the normally sampled encodings as the input of the decoder. The algorithm detail is given in the Support Information. During the sampling phase, the model only samples the encodings from the desired class region. As shown in Figure \ref{fig:samler_cmp}(a), the sampled encodings from LSATC are concentrated in several preferred regions, but those from the GMM are spread out in low score regions and those from the conditional WAE are sampled in its own cluster space. 

We compare the sampling quality of these models where the random sampling (our labelled dataset) is used as a baseline. For the overall score, the binding score and the hydrophobicty of each peptide are ranked against those of the labelled dataset. The two ranks are summed to give the overall score. The lower this score, the better the model is. The overall scores of random peptides can be considered as background ranks. For each machine learning model, 100 sequences are sampled. In the rightmost figure of Figure \ref{fig:samler_cmp}(b), the overall score distributions of the peptides sampled from all the machine learning models show an obvious shift to the lower score side when compared with that of the random model. The median and the IQR of the LSATC are 7975 and 12436, respectively, which are 6.3 times and 2.3 times smaller than those of the random model. The median overall score of LSATC is 2.8 times smaller than GMM and 4.2 times smaller than CWAE. Its IQR is 1.3 time smaller than GMM and 2.1 times than CWAE. As shown in Figure \ref{fig:samler_cmp}(b), LSATC has the lowest median and the narrowest IQR, indicating that it is the most efficient sampler among these models. When compared with the random model, LSATC samples peptides with $36\%$ lower binding score in $16$ times smaller IQR and $284\%$ lower hydrophobicity with $1.4$ times smaller IQR. Additionally, GMM sampled peptides also have low median and small IQR of binding score and hydrophobicity, indicating that high quality peptides also exist in the neighborhood of the known high quality peptides in the latent space. The performance of the CWAE model is the worst among the three models, albeit, better than the random model. The median hydrophobicity of CWAE is lower than the random model but the IQR is almost the same as the random model, indicating that sampling in the label 0 region shown in Figure \ref{fig:samler_cmp}(a) can not guarantee low hydrophobicity. This observation is reasonable because the high quality peptides (label 0) are very sparse. Modelling the distribution of the high quality regions is very difficult because such a distribution function should have many local modes while the number of modes is sparse compared to the whole region. Comparison of the performance of the LSATC, the GMM and the conditional WAE suggests that the most efficient way to sample high quality peptides is to sample in each of the high quality yet small regions.            

In Figure \ref{fig:samler_cmp}(c), we show the sampled peptides from the LSATC model. Extending amino acid residues will in general improve binding due to increased nonspecific interactions, but our model seems to generate peptide sequences to maximize specific binding. As shown in the logo plot, I, R, E, Y, K are the most frequently generated amino acids at each position and the majority of them is charged/polar residues capable of forming specific polar/ionic interactions. The results indicate that our model can learn the binding environment and then generate corresponding residues. They also corroborate that penalizing hydrophobicity in the loss function benefits the model to generate fewer nonspecific interactions. A full sample table is shown in Support Information.


\subsection{Experimental results}
\begin{figure}[h]
\centering
    \vspace{-5pt}
    \includegraphics[width=0.9\textwidth]{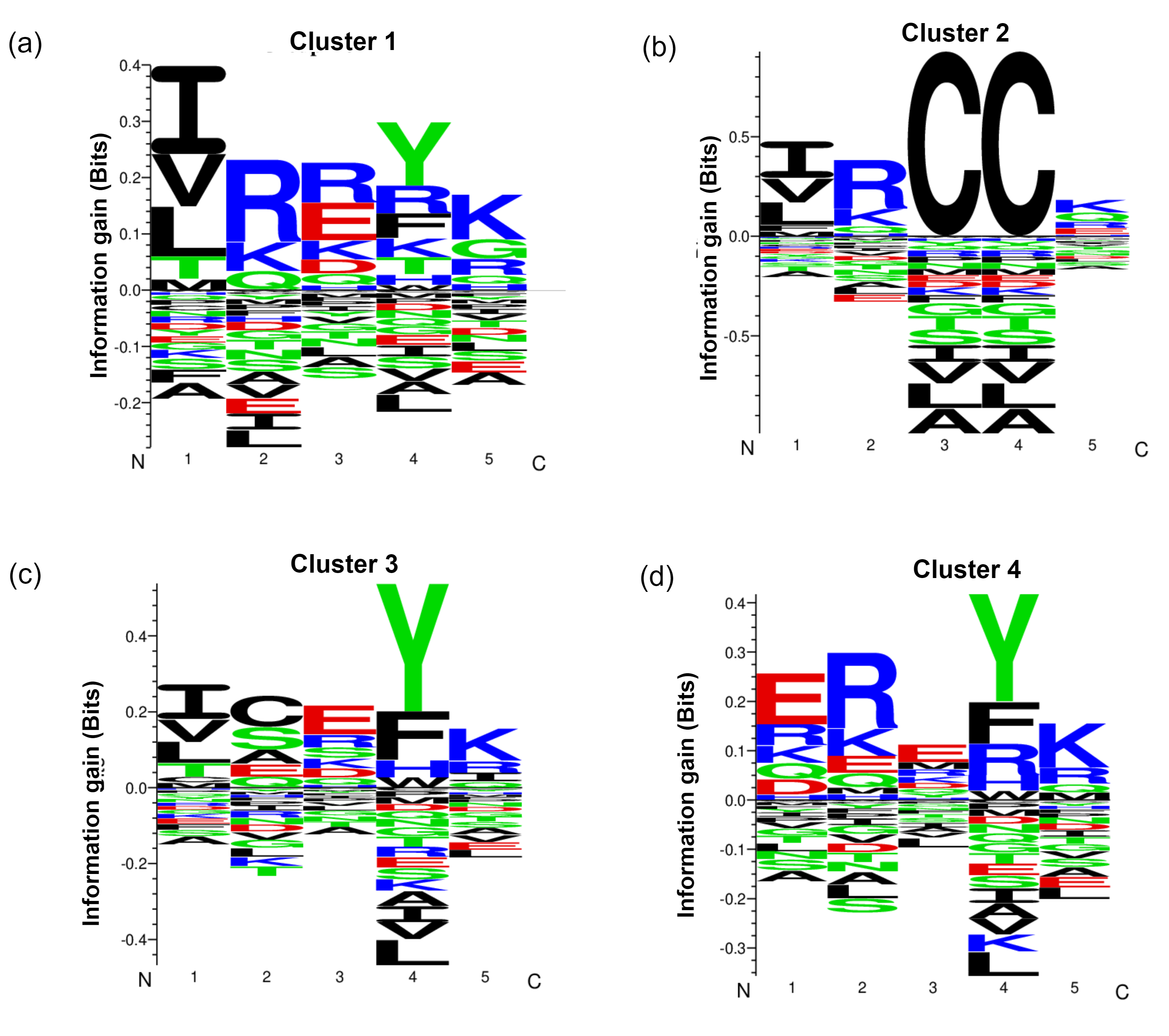}
    \caption {The logo plot of the LO matrices for the four cluster. The x-axis is the amino acid location, which has a total length of 5 in our case. The y-axis is the information gain of an amino acid at that location compared with its own background frequency. (a),(b),(c),(d) are corresponding to the logo plot of the LO matrix for cluster 1 to 4 respectively. In a logo plot, the top amino acid in each column represents the most likely amino acid at that location in this cluster. }
    \vspace{-5pt}
    \label{fig:samler_cluster}
\end{figure}

From Figure \ref{fig:samler_cmp}(b), it is evident our LSATC-generated peptide inhibitors lose differences in binding strengths and hydrophobicity. Thus, beyond selecting peptides out of the best scores, we tend to cluster peptides in terms of diversity. The strategy first selects 40 extensions with the best overall scores out of 100 LSATC sampled extensions. Second, the strategy further picks up 4 representative extensions from 4 different clusters among the elite extensions using GibbsCluster \cite{gibbscluster}. The detailed procedure is shown in Section 4.6. Figure \ref{fig:samler_cluster} shows the log-odd (LO) matrix for each of the four clusters, represented by Seq2Log \cite{seq2plot} plots. 20 natural amino acids are sized by their log-odds score at the corresponding position in the logo plot. This score can be considered as the probability of the appearance of the amino acid at this position in this cluster. For Cluster 1 to Cluster 4, each contains 16, 2, 11, and 11 sequences, and the corresponding cluster representatives are IREYK and IREFK, IRCCK, ICEYK, and EREYK. It is worth to notice the first row in the logo plot represents the theoretical amino acid for the highest likelihood appearance within the cluster and they are not always present in our sampling pools. However, all of them can be found in our LSATC-generated peptide inhibitor list except the first one. The complete 40 peptides are attached in Support Information. Overall, we integrate GibbsCluster peptide clustering method to implement a more systematic way to  select testing peptides for binding assay. 

We compare the experimental results with the parent peptide. Note that our real base peptide in experiment is "GGYPEDILDKHLQRVIL". However, it is a good practice to remove "GG" in the computational design because glycine is usually enforced to loop secondary structure, which brings flexibility and uncertainty in our binding energy evaluations. In Table \ref{tbl1:invitro}, the base peptide is GG truncated version of our original base peptide. It is clear that all the LSATC-generated peptides largely improves the binding affinity and the best peptide “ICEYKYPEDILDKHLQRVIL” has an improvement over 3 folds.

\begin{table}
    \centering
  \caption{Experimental results of the tested peptides.}
  \label{tbl1:invitro}
\scalebox{0.7}{
  \begin{tabular}{l l c c}
    \toprule
    \hline
    \textbf{Source} & \textbf{Peptide sequence}& \textbf{\emph{in vitro} IC$_{50}$ (nM)} \\
    \hline
    \textbf{LSATC-1} & IREYKYPEDILDKHLQRVIL & 51.8 $\pm$ 8.2 \\[0.3em]
    \textbf{LSATC-2}	& IRCCKYPEDILDKHLQRVIL	& 120.3 $\pm$ 2.6 \\[0.3em]
    \textbf{LSATC-3}	& ICEYKYPEDILDKHLQRVIL	& 47.8 $\pm$ 1.9 \\[0.3em]
    \textbf{LSATC-4}	& EREYKYPEDILDKHLQRVIL & 68.4 $\pm$ 10.2 \\[0.3em]    
    \textbf{base peptide}	& GGYPEDILDKHLQRVIL & 150 $\pm$ 20 \\[0.3em]
    
    \hline
  \end{tabular}
}
\end{table}

%% file: contents/3_discussion.tex
\section{Discussion}
In this study, we propose LSATC, an automated protein specific peptide design method by collecting the search trajectory of the sampler in peptides' latent space. This method opens a wide spectrum of users because all we need is an initial pose of the peptide-protein complex.

In LSATC, CMA-ES is used as an automatic encoding tuner to guide a Gaussian sampler to iteratively explore valid peptide regions with low hydrophobicity and low binding energy in the vast latent space. CMA-ES, as a gradient free method, is ideal for this peptide generation task. The reason is three fold. First, the encodings are regularized to be a Gaussian distribution in the sequence reconstruction model. This aligns with the CMA-ES assumption where the sampler is also a Gaussian. Second, CMA-ES is a gradient-free method. It is not guaranteed that any generated encoding can be decoded into a valid sequence. Thus, an important objective is to make sure that the generated encodings are valid (e.g. the decoded sequence has a length of 5). There is no gradient information about this objective w.r.t the encodings. Third, the loss w.r.t the encoding is likely to be highly non-convex. Many local minima could exist and we would like to collect the information in these local minima. It is easy for a gradient based method to be stuck in a local minimum while a gradient free method can climb over the barriers between these minima. To reduce the computational cost, a surrogate model was trained to evaluate the binding and hydrophobic properties, making the iterative search computationally feasible. We have shown that the LSATC sampler can learn a series of individual Gaussian distribution to approximate a small region at each of the local minima in the peptide latent space, which effectively avoids the encoding depletion problem. We use a peptide extending task to show that LSATC can generate b-catenin specific peptides much efficiently than the random selection and other machine learning models that globally fit the encoding distribution of high quality peptides in \emph{in silico} evaluation. 

Moreover, we propose a strategy to select reasonable number of representative peptides from LSATC proposed peptides for the final \emph{in vitro} test. In the final \emph{in vitro} test, the selected peptides are found to show improved binding affinity than the base peptide and also outperform all peptide extensions obtained from a library screening method, proving the practical usefulness of our LSATC sampler.  

As mentioned in Section 2.1, LSATC method requires around 50,000 simulated binding data w.r.t the target protein. Thus, the peptide sampling quality essentially depends on the quality of the mutation model for estimating the 3D structure of the peptide-protein complex. For short peptide generation, the mutation model is accurate and it is possible to finish the de novo design task, that is, generating complete peptides rather than peptide extensions. However, for long peptides, the accuracy of the mutation model becomes worse and using LSATC in de novo peptide design can result in poor peptides in practice. The deep learning techniques in macromolecules drug discovery is very active today. There are some works in binary classification of whether a peptide ligand and a protein receptor can bind or not by sending both sequences into a deep learning model. It is possible that in the near future a deep learning model can accurately predict a continuous binding score between a peptide and a protein. Then, the mutation model will not be necessary for the LSATC method and it will be possible for LSATC to complete de novo design of generating arbitrary length peptides in high quality.

%% file: contents/4_method.tex
\section{Method}
\subsection{Peptide binding energy calculation}
We use crystal structure of beta-catenin bound with a stapled peptide inhibitor (PDB:4DJS) as the starting structure for peptide binding energy calculation. The original peptide sequence YPEDILDKHLQRVIL was extended by 5 alanines that were enforced to adopt an alpha helical structure. The extended peptides were then superimposed onto the stapled peptide in the co-crystal structure. The poly alanine extensions were randomly mutated to any of the 20 natural amino acids, and the resulting structures were minimized by Modeller. \cite{modeller}

To search for a robust scoring function for our optimizer, we tested five binding energy calculation tools, including molecular mechanics generalized Born surface area (MM/GBSA), Rosetta FlexPepDock, flex ddG, flex ddG gam and Pyrosetta to identify the one whose predicted binding energies best correlate with in vitro IC50 values. \cite{modeller, MM/GBSA, cite5,flexddG,chaudhury2010pyrosetta}(Figure 5) MM/GBSA is a popular method for binding free energy estimation, in which a peptide-protein complex is subjected for a 10 ns molecular dynamics simulation and GMX\_MMPBSA is then used to approximate the binding energy.  Rosetta FlexPepdock is another popular method for estimating peptide-protein interaction energy. Structures of a peptide-protein complex are first sampled with a Monte Carlo method and the interface energy is then computed with the Rosetta scoring function for every sampled conformations. The mean value of 50 conformations is reported. Pyrosetta is a modified python version of Rosetta. A peptide-protein complex structure is first minimized and then subjected to interface energy calculation. In addition to the three best known binding energy calculation methods, we also tested the Flex ddG protocol, which is developed to predict binding free energy change upon mutation (interface ddG) of the peptide at the peptide-protein interface. Furthermore, the interface ddG can be improved to correlate with experimentally determined interface ddG with the generalized addition model (GAM) approach. A peptide-protein complex structure is first subjected to conformational sampling using backrub, followed by torsion minimization and side chain repacking. The mean of the interface ddG of the resulting conformation ensemble is reported as binding energies. Our results reveal that Pyrosetta has the best correlation between computed binding energy and experimental IC50 values, and Pyrosetta is thus chosen for subsequent evaluation of the ML model generated peptide sequences.

\subsection{Dataset preparation}
Two datasets were prepared: one is an unlabelled dataset for reconstruction model training, and the other is a labelled dataset for surrogate model training. The unlabelled dataset contains 500,000 randomly sampled peptides of length 5. Note that there are 3,500,000 combinations in total. We use $80\%$ of them to train our VAE and $20\%$ to test. The labelled dataset contains 50,000 randomly sampled peptide extensions. These extensions are concatenated with the base peptide. The hydrophobicity and binding energies of the extended peptides are evaluated using Biopython and Pyrosetta. It takes around 12 hours to evaluate their binding energies using multi-processing on a 16 core CPU computer. 

The binding score distribution is highly skewed. There are many high values due to steric clash between the peptide and the protein. We first remove the outliers to keep the values within 0.1 to 0.9 quantile. Then we use $\log(x+100)$ to normalize the binding scores into a small range. Note that without this data preprocessing step, the surrogate model cannot make reasonable predictions at any range. For the hydrophobicity, their values follow a normal distribution. Thus, we only standardize the values to a unit Gaussian. The distribution changes before and after the preprocessing are shown in SI Figure xx.

\subsection{Sequence reconstruction model and surrogate model}

The sequence reconstruction model consists of an embedding network, a gated recurrent unit and a multiple layer perceptron. The embedding network converts the discrete representation of individual amino acids to a continuous representation and concatenates them together to represent a peptide sequence. This ordered representation is processed by a GRU network and a MLP network to output the mean and the standard deviation of the encoding of the peptide. Then the mean of the encoding is input into another GRU network followed by a MLP network to reconstruct the peptide. 

Eq.\ref{vae_loss} shows the loss of the VAE. The first term is a reconstruction loss. It tries to match the output $x_{o}$ decoded from the encoding z and the input $x_{i}$ that encodes the z. The second term is an encoding regularizer that uses KL divergence to force the encoding to be normally distributed. In WAE, the KL divergence w.r.t p(z) is substituted by the Mean Maximum Discrepancy (MMD) measure as shown in Eq.\ref{wae_loss}. 
\begin{equation} \label{vae_loss}
    \text{vae loss}=\underbrace{-E_{q_{d}(z\vert x_{i})}\log p(x_{o}\vert z)}_{\text{reconstruction loss}}+\underbrace{D_{KL}(q_{D}(z\vert x_{i}) \| p(z))}_{\text{encoding regularizer}}
\end{equation}
\begin{equation} \label{wae_loss}
    \text{wae loss}=\underbrace{-E_{q_{d}(z\vert x_{i})}\log p(x_{o}\vert z)}_{\text{reconstruction loss}}+\underbrace{MMD(q_{D}(z\vert x_{i}), p(z))}_{\text{encoding regularizer}}
\end{equation}
The definition of MMD is shown in Eq.\ref{MMD}. It is a discrepancy measure of two distributions $q_{D}$ and $p$ after the values are transformed into a Hilbert space using some function $\phi$.
\begin{equation}\label{MMD}
\begin{split}
    \text{MMD}^{2}(q_{D},p) &=\|E_{z_{1}\sim q_{D}} [\phi(z_{1})] - E_{z_{2}\sim p} [\phi(z_{2})]\|_{H} \\
    &= E_{z_{1}\sim q_{D}}E_{z_{1}'\sim q_{D}}\langle [\phi(z_{1})],[\phi(z_{1}')]\rangle  
     -2 E_{z_{1}\sim q_{D}}E_{z_{2}\sim p} \langle[\phi(z_{1}],[\phi(z_{2}]\rangle \\ 
    &+ E_{z_{2}\sim p}E_{z_{2}'\sim p}\langle[\phi(z_{2})],[\phi(z_{2}')] \rangle
\end{split}
\end{equation}
In Eq.\ref{MMD}, the independent variables in the final form of the $MMD^{2}$ are fully defined by a valid inner product $\langle \cdot,\cdot \rangle$ known as kernel function. It is unnecessary to define the transformation function $\phi$ once we know the form of a kernel. In this study, we transform the learned encodings and normally sampled $z$s using a Gaussian kernel to calculate the MMD loss between the two distributions. In addition, we add to the unit Gaussian a weak KL penalty term directly on the learned mean and variance of the encoding. For the construction loss, we use cross-entropy (CE) loss since the input and the output are discrete. The final loss is shown in Eq.\ref{our_Loss}. 
\begin{equation}\label{our_Loss}
 \begin{split}
    loss = \underbrace{CE(x_{i},x_{o})}_{\text{reconstruction loss}}+\underbrace{MMD(q_{D}(z\vert x),N(0,1))+ 10^{-3}* D_{KL}(q_{D}(z\vert x)\|N(\mu_{z},1))}_{\text{encoding regularizer}}
 \end{split}
\end{equation}

The surrogate model structure is similar to the sequence reconstruction model. Instead of performing a reconstruction task, the output of the decoding GRU network is directed to a $7 \times 100$ image for hydrophobicity and binding energy prediction using a CNN model. The final loss is a summation of the mean squared error (MSE) loss of the two properties as shown in Eq.\ref{mse_Loss}, where the asterisk sign represents the ground truth. In practice, we tested different surrogate model structures, and the CNN model was chosen due to its best performance. In Support Information, we show the training results of all the structures that we have tested. 
\begin{equation}\label{mse_Loss}
\text{surrogate loss}=MSE(BE^{*},BE)+MSE(Hydro^{*},Hydro)
\end{equation}

\subsection{CMA-ES optimization design}
The CMA-ES is a gradient-free optimization method. It assumes a multivariate Gaussian sampler. At each step, the algorithm collects a pool of samples. The sampler mean is updated according to Eq.\ref{CMA_ES_mean}. In the equation, $\mu^{i}$ is the sampler mean at the $i^{th}$ iteration. $b$ is the learning rate. $x_{n}^{i}$ is the $n^{th}$ sampled point at the$i^{th}$ iteration. $w_{n}$ is the weight of $x_{n}$. $\sum w_{n}$ is equal to 1. $k$ is the number of the top selections measured by a fitness function. We update the mean of a sampler based on the k selected points and the previous mean. b is usually set to 1. In this case, we update the mean according to the weighted combination of the k selected points. 

\begin{equation}\label{CMA_ES_mean}
 \begin{split}
    \mu^{i+1}=\mu^{i}+b\sum^{k}_{n=1}w_{n}(x_{n}^{i+1}-\mu^{i})
 \end{split}
\end{equation}

The covariance matrix update is based on the previous covariance for a more accurate estimation. It has two components as shown in Eq xx. The first term combines the current estimated covariance with weighed previous covariances. The second term is an improvement on the first term. 
\begin{equation}\label{CMA_ES_S1}
 \begin{split}
    S^{i+1}_{1}=(1-b_{2})S^{i}+b_{2}\underbrace{(\sum_{n=1}^{k}w_{n} (\frac{x_{n}^{i+1}-\mu^{i}}{\sigma^{i}}) (\frac{x_{n}^{i+1}-\mu^{i}}{\sigma^{i}})^{T})}_{\text{current covariance estimation}}
 \end{split}
\end{equation}

As we can see in the first term, $\frac{x_{n}^{i+1}-\mu^{i}}{\sigma^{i}}$ $(\frac{x_{n}^{i+1}-\mu^{i}}{\sigma^{i}})^{T}$ loses the sign information during the update. To reinforce this information, the covariance is updated through another evolutionary path that is constructed by the mean of the previous steps rather than a single sample. Such a path is shown in Eq.\ref{CMA_ES_path}  
\begin{equation}\label{CMA_ES_path}
y^{i+1}=(1-b_{3})y^{i}+\sqrt{b_{3}(2-b_{3})w_{eff}}\frac{\mu^{i+1}-\mu^{i}}{\sigma^{i}}
\end{equation}

\begin{equation}\label{CMA_ES_S2}
 \begin{split}
    S^{i+1}_{2}=(1-b_{4})S^{i}+b_{4}\underbrace{y^{i+1}(y^{i+1})^{T}}_{\substack{\text{estimation from} \\ \text{evolution path}}}
 \end{split}
\end{equation}

Eq.\ref{CMA_ES_S2} shows the covariance updating rule from the evolutionary path. Such an update has been shown to better couple the two optimization steps\cite{cite33}. The two covariance updates are summed together to form the final update rule shown in Eq.\ref{CMA_ES_S}. 
\begin{equation}\label{CMA_ES_S}
 \begin{split}
    S^{i+1}=S^{i+1}_{1}+S^{i+1}_{2}
 \end{split}
\end{equation}

Despite many hyperparameters in the above equations, most of the parameters have optimal values \cite{cite33}. The only parameter to tune in CAM-ES is the initial sampling size. The initial mean and covariance do not affect the result so much\cite{cite34}. Although simple, CMA-ES has shown good performance in non-smooth, non-continuous functions and even noisy datasets, and is a reliable method for local optimization\cite{cite35}. 

In this work, the fitness function of the CMA-ES is the sum of hydrophobicity, the binding energy and the penalty of sampled invalid peptides. During the optimization, N peptides will be sampled from the current sampler. However, due to the large space of the encoding, it is not guaranteed that the decoded peptide sequences will have a valid token. Sometimes, the peptide length is not 5. Very occasionally, the padding token could appear before the ending token, which violates the token rule. Once these situations occur, the generated encodings will be considered as invalid. Considering that CMA-ES is a population based optimization method, we want to make sure that at each optimization step, the majority of generated samples is valid. Thus, inside each step we designed another generation loop that repeatedly samples 1000 peptides. The loop exits only when $80\%$ of the samples are valid or the loop reaches 20 iterations. Eq.\ref{CMA_ES_individual_loss} shows the fitness function $f(z_{i})$ for each of the sampled encodings $z_{i}$. 
\begin{equation}\label{CMA_ES_individual_loss}
 \begin{split}
    f(z_{i}) = w(z)(NN_{b}(z_{i})+0.5*NN_{h}(z_{i})) + w(z)I(z_{i})
 \end{split}
\end{equation}
where, \\
$NN_{b}$ and $NN_{h}$ are the surrogate model output for the binding energy and the hydrophobicity,\\
$w(z)$ is 1 if less than $80\%$ of the sampled encodings are valid. Otherwise, the value is 0. \\ 
$I(z_{i})$= 0.1*(number of the invalid encodings). 

In Eq.\ref{CMA_ES_individual_loss}, we set the weight of the hydrophobicity to 0.5 to prioritize the binding energy minimization. In our implementation, the invalid encodings are removed from the population to let the optimizer focus on valid encodings if less than $20\%$ of the samples are invalid. The detailed diagram of the optimization design is provided in SI section xx.

\subsection{Sampler selection} 
The optimization trajectory collection is necessary due to the highly non-convex optimization in the encoding space and the depletion of the encodings at a single location. To avoid collecting similar samplers, we sort them based on the loss function of the CMA-ES and select the top 500 samplers. We cluster the 500 samplers using a 10 cluster K-mean algorithm. 10 samplers that are the closest to the cluster centers are selected. The K-mean algorithm is a clustering algorithm that tries to minimize the distance between samples in individual clusters. Its objective function is shown in Eq.\ref{k_mean_loss}. Here we use Euclidean distance as the distance measure. In SI, we show the clusters of the samplers in a 2d tsne plot. It is obvious that the selected samplers are distant from each other. 

\begin{equation}
    \text{$loss_{k-mean}$}=\sum_{n=1}^{N}\sum_{k=1}^{K}I(c_{i}==k)f(z_{i},\mu_{k})
\end{equation} \label{k_mean_loss}
where, \\ 
N and K are the total number of the samplers and the K =10 is the number of the clusters.\\
I(.) is an indicator function. \\
$z_{i}$ is an sampler mean. $c_{i}$ is the cluster of the $z_{i}$.$\mu_{k}$ is the mean of the $k^{th}$ cluster.  

\subsection{Peptide selection of the experiments}
We first sample 100 peptides using LSATC and rank the peptides using the overall scores to yield top 40 peptides. We further clustered these peptide extensions with GibbsCluster\cite{gibbs_algo}. GibsCluster is an unsupervised peptide pattern discovery algorithm that simultaneously samples, clusters and aligns peptide data. The algorithm learns a m $\times$ n log-odd matrix where an element at $i^{th}$ row and $j^{th}$ column represents the information gain of the $i^{th}$ amino acid at the $j^{th}$ position compared with the background information of the $i^{th}$ amino acid. m is the 20 natural amino acids. n is 5 because we only extend 5 more residues. This information gain is proportional to the probability of an amino acid occurring at the location in this cluster. 

A peptide's representative score in that cluster is calculated via the summation of its amino acid information gain in the cluster's LO matrix. The information gain is essentially a log probability score. Thus, calculating the summation value is similar to calculate the unnormalized probability score of occurrence of the peptide in this cluster.     

\subsection{Peptide synthesis}

Peptides were manually synthesized by SPPS on Rink amide resin by using Fmoc chemistry. The in vitro IC50 of the predicted peptides against $\beta-$catenin is measured through a fluorescence polarization (FP)$\-$based competition assay. FAM$\-$labeled probe peptide (10 nM) was incubated with 50 nM GST$-\beta-$catenin in 20 mM Tris, 300 mM NaCl, pH 8.8, 0.01$\%$ Triton$\-$X100 for 1 h as reported previously. Serial dilutions of a competitor peptide were prepared in 20 mM Tris, 300 mM NaCl, pH 8.8, and 0.01$\%$ Triton-X100. After 1 h, aliquots of the equilibrated probe peptide $-\beta-$catenin solution were added to serially diluted peptide solutions and incubated for 1 h at RT. Samples were transferred into black-on-black 384-well nonbinding microplates (Greiner), and FP was measured using a Tecan M1000 Infinite plate reader. The data were analyzed using GraphPad Prism v. 8.0 and normalized to FP values corresponding to the fully bound/unbound probe.